# Space-Time Reference with an Optical Link

Paul Berceau,[b] Michael Taylor,[b,c] Joseph M. Kahn[c] and Leo Hollberg[a,b]

[a]Department of Physics
[b]Hansen Experimental Physics Laboratory & Stanford Center for Position Navigation and Time
[c]Department of Electrical Engineering
Stanford University, Stanford, CA 94305-4085
∗Email: leoh@stanford.edu

**Abstract—**We describe a method for realizing a high performance Space-Time Reference (STR) using a stable atomic clock in a precisely defined orbit and synchronizing the orbiting clock to high-accuracy atomic clocks on the ground. The synchronization would be accomplished using a two-way lasercom link between ground and space. The basic concept is to take advantage of the highest-performance cold-atom atomic clocks at national standards laboratories on the ground and to transfer that performance to an orbiting clock that has good stability and that serves as a "frequency-flywheel" over time-scales of a few hours. The two-way lasercom link would also provide precise range information and thus precise orbit determination (POD). With a well-defined orbit and a synchronized clock, the satellite could serve as a high-accuracy Space-Time Reference, providing precise time worldwide, a valuable reference frame for geodesy, and independent high-accuracy measurements of GNSS clocks.  With reasonable assumptions, a practical system would be able to deliver picosecond timing worldwide and millimeter orbit determination.

**Introduction**
An advanced high-performance Space-Time Reference (STR) could be realized using existing space-qualified subsystems, including a highly stable atomic clock that is periodically connected to high-accuracy atomic clocks on the ground via a two-way laser communication link. The satellite would be in a precisely determined orbit, and the lasercom link would also provide high-precision range data.

The system outlined here would be broadly applicable and could provide a platform for testing fundamental physics in space, such as clock and position-based tests of Einstein's Relativity, including Local Lorentz Invariance (LLI), gravitational red shift, local position invariance (LPI), frame dragging, Shapiro time delay, and two-way speed of light measurements to constrain the curvature of space-time], tests of symmetry postulates (CPT invariance); and searches for new scalar or vector fields beyond the standard model.[1,2,34,5,6,7,8,9,10,11] With a stable clocks in a well-defined low-drag orbit the General Relativistic corrections to coordinate, proper time and light propagation intervals can be calculated and tested experimentally to higher accuracy.[12,13,14,15] From the scientific perspective there is strong motivation for more precise and accurate tests of gravity; where compelling experimental evidence for Dark Matter and Dark Energy show that our fundamental understanding of gravity and "matter" is woefully lacking.

As a starting point, moving toward those more ambitious missions and science goals, this initial STR concept has three primary objectives: precise Time and precise range delivery worldwide, and precise independent measurements of GNSS clocks from above the atmosphere. In near real-time (a few seconds of averaging), we want to be able to synchronize the orbiting "flywheel" clock with state-of-the-art high-performance clocks on the ground with a precision of a picosecond.  That would include both accurate frequency and Time (epoch) and also precise measurements of the time-dependent range from the ground station to the satellite and thus Precise Orbit Determination (POD) at the millimeter level.

The concept of a space-time reference based on good clocks and a well-defined orbit is the foundation of current GNSS systems (GPS, GLONASS, BEIDOU/COMPASS, and GALILEO). What distinguishes our current approach is the specific combination of a high-stability atomic clock in a well-defined Medium Earth Orbit (MEO) and a two-way laser communication (lasercom) link that would provide much higher performance clock synchronization (ps rather than ns) between ground and space. With accurate time tagging on the two-way laser link the system would simultaneously give precise ranging for Precise Orbit Determination (POD). This approach would allow the STR to take advantage of state-of-the-art cold-atom atomic clocks at standards labs around the world. The highest-performance atomic clocks tend to be large, fragile, complex, and continually evolving and improving. Those clocks could stay on the ground and transfer the high-accuracy timing to a reliable, stable clock in orbit. This approach would allow us to take advantage of modern cold-atom atomic clocks and free-space laser links to significantly improve our knowledge of Time and position worldwide, and is achievable with a combination of current technologies and subsystems

A note on notation: to avoid confusion, we chose specific notations for "time" in this paper. The term "time-transfer" can be misleading because it is often used as a shorthand reference to the transfer or comparison of frequencies between locations but without any accurate marker of Time-epoch or a precise time-stamp in a coordinate reference frame. For clarity, we will use "Time" with a capital "T" to represent a Time stamp or Time-epoch marker in a predetermined reference frame and Time scale,[16] e.g. TAI (Time Atomic International), or UTC, (Coordinated Universal Time). We will use "time" with a lower case "t" to represent time evolution, and "Δt" to represent an interval of time.

A limitation of the highest-performance clocks today is in transferring "Time" and even accurate frequency from those clocks to other locations. Unless a practical time-transfer system can be developed, it is difficult to imagine how the advanced clocks could be put to practical commercial uses. Time (epoch) is currently realized and transferred worldwide at the 1-to-30 ns level relative to defined Time Scales by using GPS common view, or GPS carrier-phase, or via Two-Way Satellite Time and Frequency Transfer (TWSTFT) using both code and carrier phase].[17,18,19,20] With 1 ns timing uncertainty it can take weeks of signal averaging to accurately determine the frequency of the best Cs atomic fountain clocks at national standards labs that have fractional frequency uncertainties of $\approx 2 \times 10^{-16}$.[21] And if that 1 ns transfer uncertainty could be maintained, it would take over 30 years of signal averaging to compare optical clocks at the now claimed fractional uncertainties of $2 \times 10^{-18}$ – an impractical, meaningless exercise. With precise laser time- and frequency-transfer from the ground to an orbiting satellite, it would be possible to improve upon the current state-of-the-art in timing worldwide. The baseline STR system described here should be able to achieve timing accuracy and synchronization at the 1-30 ps level, a factor of about a thousand fold improvement over existing GPS and TWSTFT systems.

To transfer Time (epoch) we also require a well-defined reference frame with precise knowledge of the relative positions and motions of the clocks and users including proper relativistic corrections. The baseline STR concept presented here relies upon a stable, nearly inertial MEO orbit, for which the orbit can be determined at the 1 mm level relative to precisely known ground stations. These capabilities could also serve the important functions of enhancing the performance of existing GNSS navigation as well as precision measurements in earth sciences such as geodesy.

**Context and related systems**
*Testing General Relativity and fundamental physics with clocks in space.*
There have been several concepts and a few detailed proposals for space missions that would use advanced clocks in space to test GR, symmetry postulates, and search for new physics; those include missions like ACES, STE-QUEST, SOC, GRESE, and others.[5,9,10,100,101,102] Most of these missions require or would benefit from a high-performance method for Time transfer from the ground to space.

Two-way Doppler-cancelling microwave links for accurate frequency transfer between ground and space were pioneered nearly 40 years ago by Vessot and collaborators to measure the relativistic gravitational red shift measurement of clocks by using a hydrogen maser launched on a rocket to 10,000 km, which was compared via the two-way link to a ground-based frequency reference.[22,23] The ACES Atomic Clock Ensemble in Space[24,25,26] mission that is planned for launch in the near future will take a similar approach to measure relativistic shifts, putting further constraints on Local Lorentz Invariance (LLI) and Local Position Invariance (LPI). That mission will put a high-performance cold atom Cs atomic clock (PHARAO), a hydrogen maser, and a stable quartz crystal on the International Space Station (ISS). Those "clocks" will be compared to high-performance frequency standards on the ground using a two-way Doppler-cancelling microwave link developed in Germany that should provide exceptional timing performance.[27] Similarly, the operational approach for the QZSS satellite navigation system being developed in Japan also uses the idea of two-way microwave links to synchronize quartz or Rb frequency references on the satellites to higher-performance ground station clocks.[28] In contrast, GPS uses ground measurements of the one-way signals received from the satellites at multiple ground stations to determine clock frequencies and orbits and updates information to the satellites on roughly a daily basis.

Several groups have demonstrated high-accuracy and high-stability frequency transfer over optical fiber links with distances between laboratories ranging from 3 to over 900 km.[29,30,31,32] A number of different approaches have been studied, including one way, two way, dark fiber, and using existing telecom network hardware and protocols. Demonstrated frequency transfer instabilities have been as low as $\sigma_y(\tau) \approx 2 \times 10^{-16}/\tau$ for distances of hundreds of km and scaling with distance (L) as $\approx L^{3/2}$.[33,34,35,36] The instability achieved with optical fiber links are comparable to or better than the instability of the best atomic frequency standards today, and should not limit frequency inter-comparisons. Significant programs are underway in Europe to extend and interconnect long distance optical fiber links between national labs and institutes that have high-performance atomic clocks.[37,38] Where such fiber links are feasible and cost effective, optical fiber connections are the easiest and best approach for frequency comparisons. Optical fiber links are not generally well-suited for Time distribution, because Time (epoch) depends on independent precise knowledge of position and the transfer path in a coordinate reference frame. Fiber links could be used for relative synchronization of Time scales and Time comparisons if positions are otherwise known in the reference frame. The highest performance systems are dedicated fiber links that have been used to demonstrate the ability to compare atomic frequencies at separated locations and for physics experiments. Recent papers provide useful comparisons of time transfer by optical fibers compared to GPS carrier phase and TWSTFT.[39,47]

The potential advantages of free-space optical methods for time and frequency transfer between ground and space have long been recognized. Pioneering work for a precise optical time transfer from ground to space is done as part of the the ESA T2L2 mission.[40,41] In that case, relatively high power (30-150 mJ, 532 nm) laser pulses at low repetition rates (roughly 10 Hz to 2 kHz) are transmitted from a laser ranging ground station and retro-reflected by optical corner cubes on the JASON-II satellite. The arrival time of the return optical pulse is timed relative to a clock on the ground. An event timer on the satellite compares the pulse arrival time on the satellite to a local quartz crystal oscillator that serves as the local clock. Free-space coherent optical frequency transfer is also being studied by LNE-SYRTE and the Observatoire de la Côte d'Azur with a goal of very high-performance frequency transfer to space.[42,43] Impressive timing precision has recently been demonstrated at NIST with one-way and two-way free-space optical links with femtosecond mode-locked lasers, where excellent stability and frequency accuracy was achieved over a folded 2 km free-space path.[44,45,46]

In principle, these free-space links could be configured to provide frequency, range, and time differences for transmitters (Tr) and receivers (Rx) within line of sight. A recent special issue of *Comptes Rendus Physique* on the measurement of time provides an excellent summary of the latest developments and current state of the art in atomic frequency standards, time/frequency transfer, Time Scales and proposed space clock missions.[47] To maximize utility and enable worldwide distribution, it would be a tremendous advantage to have the Time, frequency and range information come from a space-Time reference in a precisely known orbit, as we are proposing.

**Reference frames**
For practical use in accurate space-time measurements on the rotating earth, the preferred reference frame is usually an Earth Centered Earth Fixed (ECEF) frame, as chosen for GNSS systems and time scales, such as TAI, UTC, GPS time. In all frames careful attention must be paid to relativistic effects of moving clocks and earth's gravity.[12,14,15,48,49] As pointed out in these references and elsewhere, at the levels of timing precision (ps) and range uncertainty (mm) that are of interest for STR the relativistic corrections will be relatively large and will require calculations to higher order in velocity ($v/c$) and gravitational potential ($\Delta U/c^2$) than in current GNSS systems.

Accurate knowledge of position on earth comes primarily from accurate geodetic surveys combined with precise GNSS measurements to create the best model for an International Terrestrial Reference Frame (ITRF).[50,51] The current TRFs are able to determine relative base station positions on the earth to about 1 mm precision at 100 km distances, and relative to the center of earth with an uncertainty of a few millimeters.

**STR system concept**
The basic concept of a space-time reference provided by stable atomic clocks in known orbits is the very foundation of the GNSS systems. The modern variant of those ideas that we are proposing would take advantage of state-of-the-art atomic clocks and laser technologies. The high performance cold-atom atomic clocks would stay on the ground at national standards laboratories (or elsewhere), and two-way Doppler-canceling laser links would be used to transfer that performance to a stable atomic clock in space that would orbit the earth with a period of several hours. The high-performance reference signals would be uplinked via optical ground terminals to correct and steer the Time and frequency of the orbiting clock.

The major system components of the STR are illustrated in Fig. 1. In this example we assume a high-accuracy optical atomic frequency standard on the ground that stabilizes a femtosecond-Optical Frequency Comb (OFC).[52] The train of phase-coherent ultra-short pulses emitted from the stabilized OFC generates the accurate and stable RF/microwave timing reference for a two-way Doppler-cancelling lasercom data link between ground and space. The precise timing information from the optical pulses is transferred to the clock generator for the data stream of the two-way lasercom link between the ground and the satellite. This approach does not require sending the femtosecond pulses directly on the laser link, but rather just precise and accurate knowledge of the timing (phase) of the data stream.

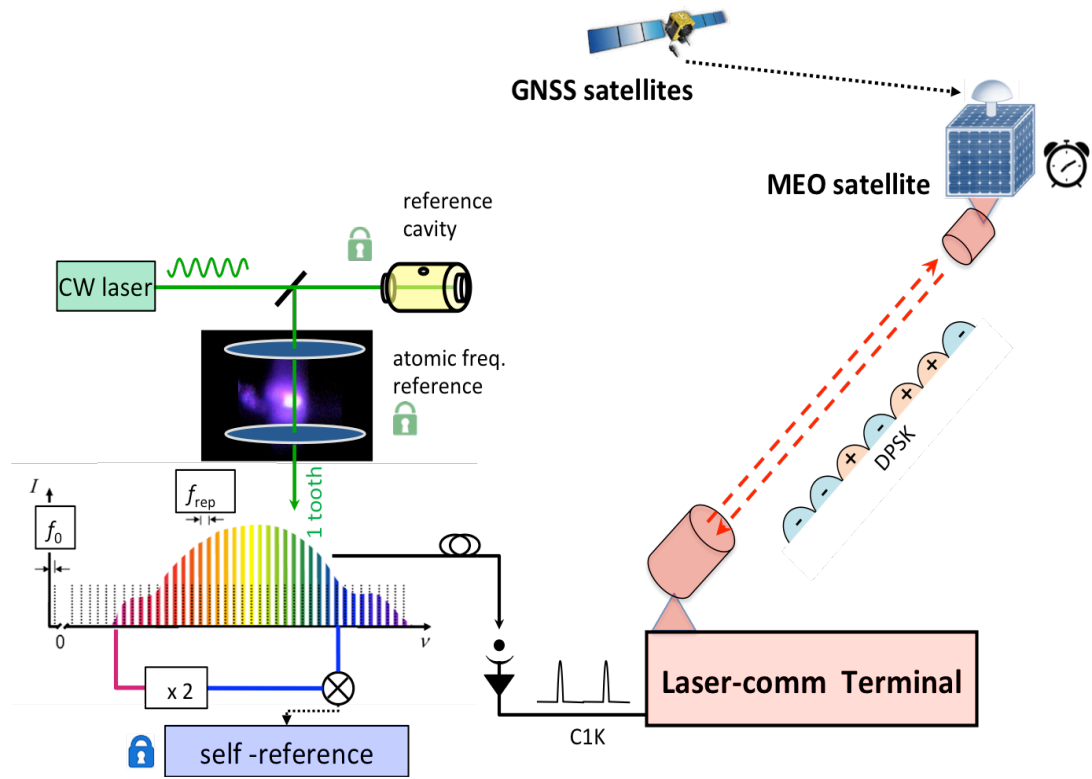

***Fig. 1.*** *Diagram of a high-performance optical time-transfer system that links a ground station with a satellite. This realization assumes an optical atomic clock on the ground and a microwave atomic clock in orbit. In the ground station, a stable CW laser is locked to both an ultra-stable Fabry-Perot cavity (on short times) and a narrow optical atomic transition (on longer times, as in a typical optical lattice or single ion clock). That laser stabilizes a self-referenced optical frequency comb that coherently down-converts the optical reference to a microwave frequency that serves as the timing reference for the two-way lasercom link. The satellite carries a stable atomic clock that serves as frequency-flywheel and provides the clock for the data stream for the lasercom downlink. The two-way Doppler-cancelling link measures the frequency difference, the time offset, and range between the satellite clock and the ground clock.*

### Lasercom Link

Several approaches could be used to decode the precise timing information, which would include frequency difference, relative phase, and Time epoch, from a data-modulated optical signal. The lasercom system could downconvert the optical signal using a local oscillator laser and coherently track the optical carrier[53,42] or perform direct detection of the optical signal and coherently track the microwave carrier underlying the data modulation.[54,55] For simplicity and guided by plans for future space lasercom data links in the U.S., a baseline system for meeting our performance goals would be a two-way 1550 nm link using Return-to-Zero Differential Phase-Shift Keying (RZ-DPSK) modulation at about one Gb/s and using direct detection. The uplink and downlink data rates need not to be equal, but precise timing and phase referencing is required on both ends of the link. Doppler shifts due to satellite motion and frequency offsets between the high-performance ground clock and the stable satellite "flywheel" atomic clock will be subtracted and recorded by proven methods of coherent RF-microwave synthesis. Timing and ranging information will come from the recorded time-series of relative phase measurements between the received clock component of the data modulation and the microwave local oscillator. The two series of phase measurements (indexed "i" for ground and "k" for space) can be written as:

$$\Delta\varphi_{gi} = \varphi_g(t_{gi}) - \varphi_{rg}(t_{gi}) \quad \text{with} \quad \varphi_{rg}(t_{gi}) = \varphi_{ts}(t_{gi} - \Delta t_{sgi})$$

$$\Delta\varphi_{sk} = \varphi_s(t_{sk}) - \varphi_{rs}(t_{sk}) \quad \text{with} \quad \varphi_{rs}(t_{sk}) = \varphi_{tg}(t_{sk} - \Delta t_{gsk})$$

where $\phi_g(t_{gi})$ represents the phase of the ground clock "g" and $\phi_{rg}(t_{gi})$ is the phase of the received ("r") data clock "r" on the ground at a local proper time $t_{gi}$. Likewise, the $k^{th}$ relative phase measurement in space is $\Delta\phi_{sk}(t_{sk})$ at local proper-time $t_{sk}$ in the satellite. The propagation time intervals between the satellite and ground and the ground and the satellite, are $\Delta t_{sgi}$ and $\Delta t_{gsk}$ respectively. The propagation delays are not identical because of satellite motion and atmospheric asymmetry and turbulence. Any relative clock offsets are contained within $t_{gi}$ and $t_{sk}$.

By including additional time-stamp information as packets on the lasercom data stream it will be straightforward to remove the integer phase ambiguity on the lasercom clock signal, which is assumed to be about 1 GHz for a RZ-DPSK system.

### *Atomic clock references*
*Current state-of-the-art ground clocks (e.g., at Standards Labs)*
Current state-of-the-art microwave atomic frequency standards based on laser-cooled atoms (Cs, Rb fountains) achieve an instability of fractional frequency, $y=\Delta f/f$, of $\sigma_y(\tau) \approx 1\times10^{-13}/\tau^{-1/2}$ or somewhat better, expressed as an Allan deviation with averaging time $\tau$ in seconds. With time averaging that instability generally decreases as $\tau^{-1/2}$ (as white frequency noise) for $\tau$ of approximately 1 to $10^5$ s.[56,57] On longer time scales we can examine "accuracy evaluations" and monthly reports of primary frequency standard comparisons as part of TAI clock comparisons.[58,59] Those show fractional frequency "accuracy" of about $2\times10^{-16}$, and that TAI is maintained at the best sites with uncertainties of a few nanoseconds. Measured locally, the primary microwave frequency standards can support fractional time interval, $x=\Delta t/t$, uncertainties of about $\sigma_x(\tau) = 1\times10^{-13}\tau^{1/2}$, corresponding to $\Delta t \approx 30$ ps at one day.

The stability and "accuracy" of optical atomic frequency standards are roughly a factor of a hundredfold better than those of the microwave clocks and have demonstrated $\sigma_y(\tau) \approx 2\times10^{-16}/\tau^{-1/2}$, which would correspond to achievable local timing uncertainties of $\sigma_x(\tau) \approx 1\times10^{-16}\tau^{1/2}$, or roughly 300 fs at one day.[60,61,62,63] However, and for many reasons, the optical atomic frequency standards are not currently being used to maintain actual Time-Scales or for distribution of Time or frequency to general users. The advanced optical atomic frequency standards tend to be research projects focusing on improved performance for basic science and fundamental physics experiments rather than on Time keeping, or Time distribution from a specified reference frame.

*Space Clocks*
The proposed STR concept requires a high-stability atomic frequency standard that can operate continuously and reliably in orbit for many years. Developing and validating the performance of space-qualified atomic frequency references is a major effort that requires significant time and cost. Fortunately, we can take advantage of decades of atomic clock development for the global navigation systems GPS, GLONASS, BEIDOU, and GALILEO. Some of the GNSS atomic frequency references have demonstrated operational lifetimes of exceeding 10 years in orbit, and given the simplicity of the standards the performance of the current generation of space atomic clocks is rather impressive. In the near future we anticipate that the high-performance PHARAO cold-Cs frequency standard will be demonstrated on the International Space Station as part of the ESA-ACES mission.[26] In addition, the compact Hg+ frequency standard developed at JPL shows great promise for future space missions.[64,65] The frequency instabilities reported for some of those space-qualified atomic frequency references are plotted in Figure 2.

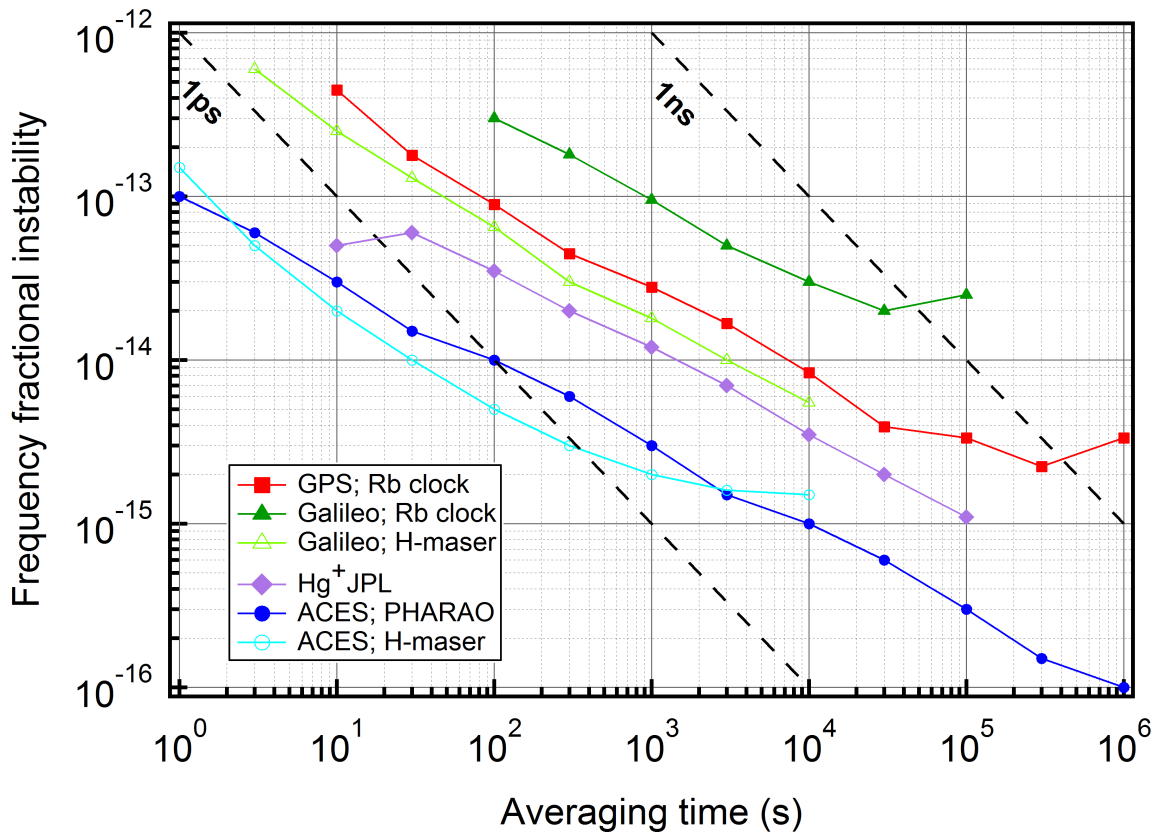

***Fig. 2.*** Fractional frequency instabilities of some space-qualified atomic frequency references as a function of averaging time τ, with data taken from published literature[66,67,68]. All of these are microwave frequency references. The red squares represent GPS rubidium clocks, the green triangles are Galileo rubidium, open green triangles are the Galileo hydrogen-maser, purple diamonds are the Hg+, solid blue circles are ACES-PHARAO cold cesium clock, and open blue circles are the ACES hydrogen maser. Two dotted lines indicate time interval uncertainties of 1 ps and 1 ns.

### STR satellite and Precise Orbit Determination (POD)

To achieve the STR performance goals for Time (epoch) and accurate ranging for geodesy we require a well-defined and stable orbit that has minimal external perturbations, and that has good visibility to key ground stations. The choice of orbit is a compromise between several driving factors:

- An orbit that is well above the troposphere can minimize atmospheric drag and associated disturbances that vary with time. For MEO satellites the dominant residual disturbances are radiation pressure and solar wind, with an estimated acceleration of $\approx 10^{-9}$ m/s$^2$ for a 1 m$^3$, 1000 kg satellite).[69]
- The orbit should be low enough to have good visibility of GNSS satellites above.
- The orbit should be high enough to be above most of the free-electron content of ionosphere, thus minimizing atmospheric group delays of troposphere and ionosphere for optimal unperturbed measurements of the microwave GNSS signals received at the STR satellite.
- The orbit's ground track should have the satellite visible to national standards labs and key ground stations several times each day for reasonable measurement times, and should achieve near worldwide coverage.

- The orbit should be high enough to allow simultaneous two-way optical lasercom links between continents, for example between Standards labs, US to Europe or Asia.

The orbit will be optimized considering these factors and using careful modeling of external disturbances, including earth shadow, solar wind, sun-moon perturbations, and magnetic effects.

A satellite at an altitude of a MEO orbit has several advantages relative to LEO orbits that are plagued by atmospheric drag and short observation times from ground stations. In MEO, the high performance GNSS receivers would also be above most of the free electron density of the ionosphere, which would allow high accuracy orbit determination from GNSS signals and the capability of monitoring the atomic clocks on GNSS satellites from a stable platform above the troposphere and most of the ionosphere, while referenced to the best atomic clocks on the ground. An example orbit that can meet most of our requirements is illustrated in Figure 3.

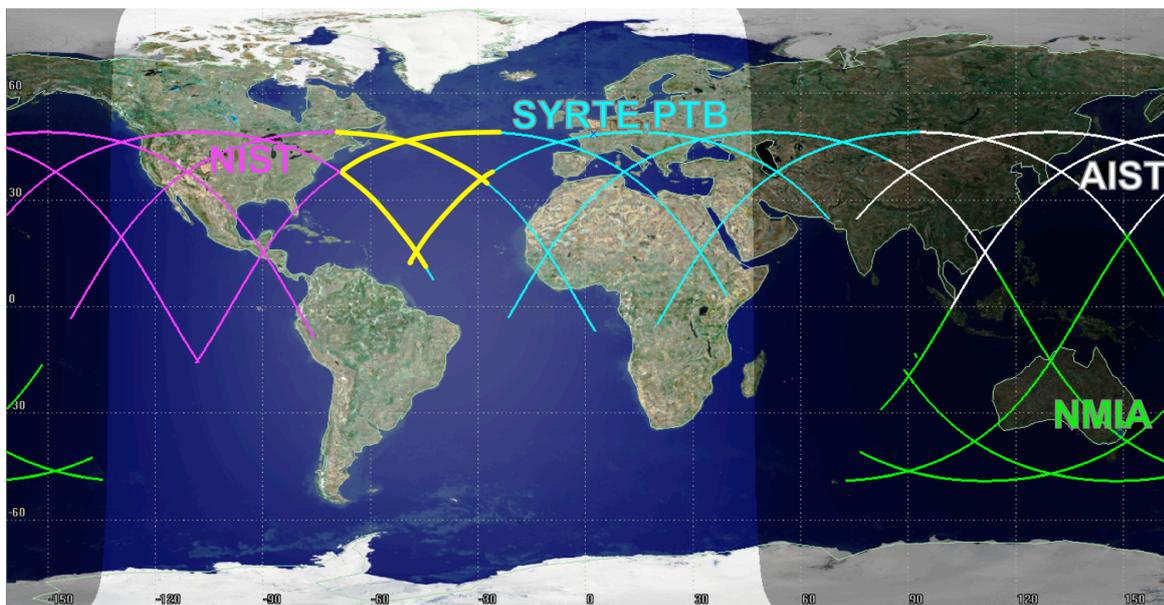

*Fig. 3. Satellite ground tracks for two days of a 5-hour MEO orbit, with inclination angle $49^0$, showing ranges of satellite visibility with elevation angles > 15 degrees above the horizon. The different colors indicate visibility for four ground sites at standards labs; NIST (USA) in purple, Europe in light blue (LNE-SYRTE France, PTB Germany, NPL UK, INRIM Italy), AIST(Japan) in white and NMIA (Australia) in green. For most of the orbits there are regions along the trajectories that allow simultaneous common-view between two or more of the ground sites (e.g., the yellow sections show the overlapping visibility between Europe and NIST).*

For high-accuracy ranging and POD the physical structure of the STR satellite should be compact and dense, with a precise knowledge of the center-of-mass relative to the lasercom optical aperture. The space terminal aperture can be relatively small, with a diameter of order 10 cm, but it must be precisely steered to connect to specific ground stations, as has been demonstrated in recent space lasercom missions. As feasible, we want to minimize moving masses and unbalanced torques and avoid steered solar panels and any asymmetric mass redistribution resulting from thruster usage, etc. In addition, we need well-calibrated and precisely controlled thermo-mechanical and electronic systems so that optical and electronic time delays through the entire system can be known and controlled at a level below 1 ps level.

With high-performance space GNSS receivers, careful orbit modeling, and advanced post-processing methods, it is now possible to determine satellite orbits post-facto at the roughly 5 cm level.[70,71] Figure 4 illustrates how a MEO orbit could enable simultaneous laser links between continents.

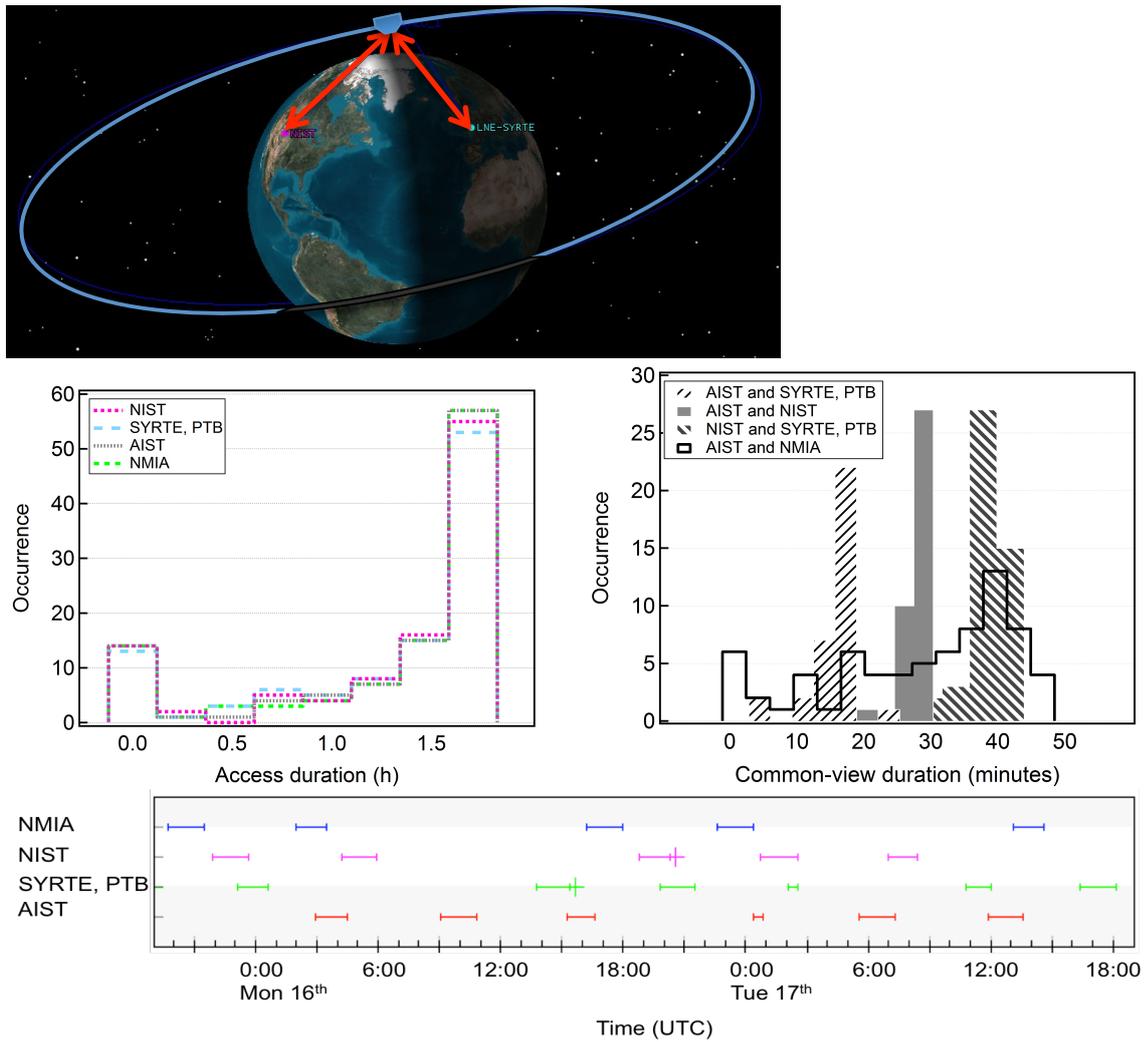

Fig. 4. The top left shows the number of occurrences in one month that the STR satellite would be visible above key national standards laboratories as a function of the duration of visibility binned into half hour intervals. The bottom shows the time series of windows of visibility for four labs over two days. Typical visibility would be about 3 hours/day.  For this example we assume a MEO orbit period of about 5 hours (inclination angle 49 deg.) and an optical link that is usable for angles greater than 15 deg above the horizon. The standards labs[72] indicated are: NIST (in red), LNE-SYRTE and PTB (in blue), AIST(in black) and NMIA (in green). All have high-accuracy atomic clocks and Time scales. On the top right is the number of occurrences in a month and the duration that two labs would have a simultaneous common-view of the STR. Noteworthy are pairs that are not likely to be connected by optical fibers (NIST-Europe, NIST-Asia, Asia-Europe, Asia-NMIA, etc).

Lasers and optics have several advantages for precise time and range relative to microwave systems.  The shorter wavelengths make it easier to determine positions accurately relative to Tr/Rx antenna or telescope apertures; in addition, lasers can be directed with low-divergence spot beams for high effective gain. Optical signals can accommodate high data bandwidths that provide leverage for precise timing and range.[73]  Microwave antenna, or optical terminal, phase-center variations with angle and their relationship to the satellite center of mass must be modeled and corrected for POD.[106] In that regard optical systems have advantages due to the shorter wavelengths. Laser links do have significant disadvantages when there is strong atmospheric scattering such as clouds.  A combination of microwave and optical links would give the best performance and most general robust solution for precise PNT.

*A baseline STR satellite would include:*
- A stable, reliable atomic frequency reference serving an orbiting clock; plus a backup.
- A high accuracy Time/phase measurement system.
- A two-way Doppler-cancelling lasercom link, compatible with space lasercom data systems.
- High-accuracy GNSS receiver with unobstructed view upward to receive multiple GNSS signals.
- Precise satellite pointing and positioning, by taking advantage of inertial navigation sensors, star trackers, GNSS and augmentations.

*Optional high-value add-ons*
- Two-way Doppler-cancelling microwave link, such as that developed by Timetech for the ESA-ACES mission.[74] That would provide complementary information on range and time transfer and additional capability during cloud cover and for atmospheric science.
- Drag-free test mass and control system[75] to reduce external disturbances and provide a much-improved inertial reference frame. This would be required for some tests of fundamental physics and gravity.
- DORIS and VLBI receivers could provide additional information for POD.

**Lasercom link for STR**

Groundbreaking work has been done in developing space lasercom links for high-speed data transfer from satellites to ground and also between satellites. To enhance the near-term feasibility of our STR concept, we want to leverage existing technologies and build toward compatibility with the next generation of space lasercom links. Doing so will reduce costs and development time, and also make it possible to add precise Time, frequency, and range information to other future space missions that might incorporate lasercom links for high-speed data transmission. The additional capability of precise Time and range would not require much additional hardware and could bring powerful new capabilities to missions.

We are encouraged by the impressive results achieved by the space lasercom community, and we would like to leverage what they have learned for our STR. An example is the recent Lunar Laser Communication Demonstration (LLCD), by NASA and MIT Lincoln Laboratory that demonstrated a duplex lasercom link between ground and the moon that achieved 622 Mb/s data downlink and only required a 0.5 W laser, 30 kg payload and 140 W total power on the LADEE satellite orbiting the moon.[76] LLCD also achieved range measurements between the ground and the moon approaching 1 cm range precision. That was achieved without a high-stability atomic clock on the satellite.[77] The ESA-DLR coherent lasercom links have demonstrated high data rates up to 5.6 Gb/s between satellites and between satellites and 1.8 Gb/s rates to the ground.[78,79] Other space lasercom demonstrations include the NASA-OPALS mission[80] to the International Space Station and space lasercom systems in Japan.[81] In the near future, NASA's LCRD[55] mission is designed to demonstration a data rate of about 2.88 Gb/s between a GEO satellite and ground stations. Considerable effort is now going toward developing systems for lasercom links to deep space, i.e., to Mars and beyond.[82]

A STR system based on a lasercom link would make use of the timing information that is essential to the function of the communications system. For the data links to function properly, the root-mean-square timing jitter must not exceed a few percent of the bit period,[83] which bodes well for precise time transfer. At both end of the link, the STR system would precisely measure the phase of the clock of the received data stream relative to the local clock. LLCD used a 200 ps slot clock, which required synchronization to a few tens of picoseconds. This precise synchronization enabled comparison of the uplink and downlink symbol clock phases that led to the impressive 1 cm range precision.[77] Similar levels of synchronization will be required for LCRD which plans to use a symbol period of about 260 ps. With phase coherent-detection of the data stream clock phase, the jitter per symbol averages down, providing sufficient signal-to-noise to make relative phase measurements with less than 1 ps uncertainty over averaging times of a few seconds

To illustrate the approach for our proposed two-way Doppler-cancelling laser-link for Time and range, we sketch out an example system that could meet our performance goals while taking advantage of considerable previous and current work for future space missions. As mentionied above, such a system could use 1550 nm lasers and components, with both data and timing information encoded on the lasers using RZ-DPSK. The high-speed data would primarily serve mission goals other than time and range transfer. To enable those additional functions, highly-stable and accurate atomic frequency references would provide the clock reference for the data link, and Time-stamping would be inserted in the data stream to identify the Time-epoch and enable precise range determination. A Time-stamp would indicate that a specific clock transition represents an on-Time epoch mark when transmitted. Both receivers would measure the relative phases of the received clock and Time-stamps relative to the local clock phase and Time. The net result would be a time series of relative (received-vs.-local) clock phase measurements on the ground and in the satellite with local Time-stamps.

To estimate the timing performance achievable with space lasercom links, we start with system parameters that are representative of the LLCD system and the planned LCRD mission. We assume the downlink and uplink both employ RZ-DPSK at bit rates of about 1 Gb/s. The downlink transmits using a 1550 nm laser, an optical power of 1 W, a beam divergence of 30 µrad, and the ground terminal uses a 40 cm receive aperture and state-of-the-art photo detectors, (either optically preamplified p-i-n photodiode receiver or photon counting receiver). For now, in this example we ignore the degrading effects of atmospheric turbulence. Guided by analysis and results from previous space lasercom missions[84] we make a rough estimate of the downlink error budget:

| | |
|---|---|
| Laser power transmitted (peak, RZ-DPSK) | 30 dBm |
| Transmit divergence FWHM [θ]= 30 µrad | |
| Tracking pointing errors | -3 dB |
| Atmospheric scattering & absorption | -2 dB |
| Receive telescope to optical fiber | -10 dB |
| Range loss for a 40 cm receive aperture vertical, orbit altitude 8,000 km | - 53 dB |
| Net link loss | - 68 dB |
| Received power | - 38 dBm |
| Power required for RZ-DPSK 1 Gb/s link | - 55 dBm |
| Link Margin = | 17 dB |

To estimate the performance of the uplink, we assume a wavelength of 1560 nm, a transmit power of 10 W, and a beam divergence angle of 30 µrad, and assume the satellite receiver has an aperture of 13 cm and uses an optically preamplified p-i-n photodiode receiver, resulting in the following estimated uplink error budget:

| | |
|---|---|
| Laser power transmitted (peak, RZ-DPSK) | 40 dBm |
| Transmit divergence FWHM [θ]= 30 µrad | |
| Tracking pointing errors | -3 dB |
| Atmospheric scattering & absorption | -2 dB |
| Receive telescope to optical fiber | -10 dB |
| Range loss for 10 cm receive aperture | - 65 dB |
| vertical, orbit altitude 8,000 km | |
| Net link loss | - 80 dB |
| Received power | - 40 dBm |
| Power required for RZ-DPSK 1 Gb/s link | - 55 dBm |
| Link Margin = | 15 dB |

An attractive feature of this STR approach – using high-performance clocks on the ground and the high signal-to-noise ratio provided by a two-way lasercom link – is that precise timing and ranging measurements should be possible over measurement times of just a few seconds without significant post-processing and analysis, and could be delivered worldwide.

To deliver Time as well as frequency and range the time delays on the optical link due to the atmosphere must be determined within the required delay uncertainty (mission-dependent). Considerable work has been done in evaluating optical delays for laser ranging to satellites and to the moon. Advanced models of atmospheric optical group delay have been developed and evaluated for wavelengths relevant for satellite laser ranging systems, such as ILRS systems at 532 nm and 1060 nm. Those models are able to provide delay uncertainties equivalent to range uncertainties of about 1 mm (corresponding to Δt uncertainties of about 3ps).[85,86] The atmospheric delay at 1550 nm will be somewhat less than at the shorter wavelengths. In addition to those models, space lasercom systems may use multiple wavelengths to reduce interference between the transmit and receive channels, to increase data rates by wavelength-division multiplexing, or for other mission requirements.[73,87] In favorable cases, multiple wavelengths could provide additional information about the net atmospheric group delay when combined with realistic models of the atmosphere and estimates of water vapor and $CO_2$ content. The actual improvement in range determination that would be achievable with multi-wavelength laser links to space remains to be proven. Multi-wavelength methods are commonly used for the highest-accuracy distance metrology through the atmosphere on the ground.[88]

Even without a good clock in orbit and other input for POD, the LLCD showed ranging precision between the ground station on earth and the LADEE satellite orbiting the moon at the few cm level.[89] Other estimates indicate that using lasercom systems for ranging at the mm level is feasible even for deep-space missions.[90,91,92]

**Feasibility**
At this juncture in time, all of the technologies and subsystems required for a baseline STR system are available, including advanced microwave and optical atomic clocks plus femtosecond frequency combs that can remain on the on the ground, space lasercom links, space-qualified flywheel atomic clocks, and precise electronic timing systems. To reduce development time and cost, we are guiding our baseline STR design to be compatible with technologies that are currently available, or soon to be deployed space-qualified subsystems that could realistically be assembled for a space mission in the next few years. STR would provide valuable data and new capabilities, and enable new science missions as well as compelling applications.

The simplicity of the STR design and compatibility with space lasercom data systems would enable the straight-forward addition of precise timing and range information to other space missions that might use lasercom links for high-speed data transfer. This alone could bring significant value, new capabilities, and new science to space missions. All of this can be achieved with ps timing and mm ranging using existing space-qualified technologies.

**Future upgrades to STR**
Higher levels of timing precision are certainly feasible and foreseeable for the future by putting advanced cold-atom optical clocks in space along with femtosecond optical frequency combs and perhaps coherent optical links for the time/frequency transfer. It would be straightforward to upgrade the STR approach outlined here if and when higher-performance space clocks become available. Considerable effort and investment will be required to engineer optical atomic clocks for robust reliable operation in space, but that could be done if there are sufficiently compelling reasons. Beyond fundamental physics experiments, practical and economically viable applications of 15 to 18 digits of frequency accuracy or stability have yet to be identified. Currently, high-performance optical atomic clocks are not ready for space, but some preliminary research is working toward that direction, e.g., the Space Optical Clock (SOC) project in Europe.[93,94] In contrast to the current status of optical clocks for space, there have been two recent demonstrations showing impressive results for a mode-locked femtosecond laser operating in orbit by Korea,[95] and a frequency stabilized fs optical frequency comb on a sounding rocket in Germany.[96]

A potentially valuable upgrade to STR would be to add an inertial test mass to the satellite to further mitigate non-gravitational disturbances and provide higher-precision orbit determination. STR could also be operated as a drag-free satellite[97] as used for GPB[98] and GOCE and being developed for other missions, e.g., LISA PATHFINDER and related.[99,75] However, long-term operation as a drag-free satellite would bring requirements for thrusters, fuel, and increased system complexity

**STR: enabling science missions and applications**
*Reference frames*
STR in a nearly inertial MEO orbit, with a stable atomic clock and a laser two-way timing link would provide an exceptional reference frame for geodesy, earth sciences and navigation systems. In the longer term, putting the same technologies in a highly elliptic orbit or free-flying closer to the sun would enable accurate measurements to test the foundations of General Relativity and our understanding of space-time to higher orders in velocity (v/c) and with larger gravitational potential differences than can be done near the earth. That approach has been proposed for several space science missions, e.g., EGE, SAGAS, STE-QUEST, GRESE, LARASE, etc.).[10,100,101,102,103]

Existing GNSS systems combined with advanced analytical methods do amazingly well in monitoring ground motion and elevation changes at the mm/yr level.[104,105] Such measurements are critical to earth sciences and hydrology. While developing these STR concepts, we learned of related ideas for improved satellite references for earth sciences that are motivated by needs for improved terrestrial references frames, more accurate measurements of sea level height, and earth gravity mapping. Some proposed missions, including GETRIS and GRASP, require new satellites in well-defined orbits whose position is determined by existing GNSS, DORIS, and VLBI systems combined with modeling.[106,107] A recent variant of GRASP proposes the addition of a high-sensitivity accelerometer, such as that used for GOCE, for inertial corrections to the orbit.[108] Those mission concepts do not propose to use a laser link for precise Timing and range determination nor a high performance atomic clock in orbit, but they would use established methods for POD, such as GNSS, DORIS and VLBI. The needs for higher-accuracy geodesy for earth sciences and the environment are real and compelling. Perhaps there could be some synergism between the proposed missions for geodesy and gravity and our STR concepts.

*Timing*
The National Standards labs, including NIST, PTB, LNE-SYRTE, and NMIJ, and some GNSS-geodetic reference sites maintain Time-Scales within their laboratories. The primary examples are $UTC_{SL}$, where SL designates specific lab, and Time Atomic International (TAI) compiled by international agreement via the BIPM. Synchronization of time scales is achieved at the few ns level, limited by time transfer uncertainties.[47] Advanced high-performance cold-atom atomic clocks are now working extraordinarily well, but have not found a user base outside the fundamental physics community. To a large degree that is due to the lack of compelling real-world applications that require stability at $10^{-15}\tau^{-1/2}$ or 15 to 18 digits of frequency accuracy. In general, the high-performance clocks and Time Scales are not accessible for potential users, and their use has been limited to a small number of local in-laboratory fundamental science experiments, such as tests of relativity, searches for time variations of fundamental constants, or searches for scalar or vector fields that may couple to clocks or gravity. To date, the results of those experiments agree with predictions of standard theories and have not identified new physics. We also lack convenient methods to take advantage of those high-performance atomic frequency references and Time Scales.

If a system like our proposed STR were available and could transfer Time worldwide with higher accuracy than available from GNSS systems, then international atomic time scales such as TAI could take advantage of the higher stability and accuracy that is now available from cold-atom atomic clocks, both microwave and optical. Generally the Standards Labs only distribute Time and frequency for widespread use with relatively low-accuracy signals by VLF radio signals or by internet time signals. More accurate Time and frequency can be transferred at considerable time and effort by setting up TWSTFT systems, as was done for recent measurements of the speed of neutrinos, which required a major effort and ultimately achieved Time uncertainties at the 500ps level.[109] The STR approach could be used to transfer Time (epoch) and frequency from high-performance clocks at National Standards labs to relatively small transportable ground terminals, with roughly. 30 cm receive telescopes essentially anywhere on earth. The national and international Time Scales (e.g. $UTC_{NIST}$, TAI) could be accessible outside of the laboratories. With two optical terminals on the STR satellite a direct two-way optical link could be established even between continents, as shown in Fig.4.

There are also a few commercial applications that would benefit from higher-precision timing than is currently available. Examples include ultrafast trading of stocks and financial transaction, where the challenges are not clock performance but rather widespread distribution, continuous availability, and methods to validate and encrypt accurate time-stamps in a well-defined reference frame. Other systems require precise synchronization of events or networks, or distribution of phase-coherence, such as interferometric coherent imaging, multi-static radar or ladar, some types of secure information transmission, mitigation of jamming of communication or navigation signals, precise ranging or position determination, and formation-flying of moving platforms, etc. Those types of systems would benefit significantly from synchronized oscillators and distribution of precise atomic time and range information.

*STR value to GNSS*
With new GNSS systems coming on line, we will soon have more than 100 stable atomic clocks in space providing navigation and timing signals. In a precisely known MEO orbit the STR satellite would provide high-accuracy measurements of the multitude of GNSS signals (precise clocks and orbits) that would be unperturbed by the troposphere and the majority of the free-electron content of the ionosphere. That could give valuable information to the GNSS services, the geodetic, and time-and-frequency communities, and for terrestrial reference frames. The STR GNSS clocks could be measured relative to the STR clock and via the laser link to the highest-accuracy clocks located on the ground. This approach would help to separate GNSS clock and orbit information from atmospheric time delays and scintillation and may provide value to the GNSS systems or precision PNT solutions.

The STR satellite would use a steerable lasercom terminal to connect to ground stations. The downlink laser could illuminate a relatively small area on the ground with a spot beam of about a hundred meters to a kilometer and could provide precise timing and range information to users that might not have access to GNSS signals, and could provide a secure high-speed data link. The STR approach will not serve as a universal navigation or positioning system, but rather a system that could augment existing GNSS or other PNT systems with higher-accuracy timing signals, clock synchronization and range information, and could provide connections to ground sites a few times per day.

**Summary**
The proposed STR approach would allow creation of a high-performance time reference in an unperturbed orbit and well-defined coordinate system, which will provide very compelling performance for future science missions and practical applications to PNT and geosciences and accessible worldwide.  With precise and accurate space-time measurements that would be analyzed and evaluated in the context of General Relativity, with an emphasis on looking for anything inconsistent with principles of LLI, LPI and gravitational red shift.  STR would build practical technologies enabling missions deeper into space for more stringent test of relativity and searches for new physics beyond the Standard Model, such as constraints on possible scalar or vector fields that couple to atoms or gravity, or CPT invariance) and potentially searches for dark matter by gravitational perturbations or via couplings to atomic energy levels.

In its simplest form, STR could be built from existing space-qualified or -qualifiable subsystems and would bring significant value and new capabilities to Time/frequency transfer worldwide, to precision geodesy and terrestrial reference frames, to earth-environmental science and to navigation systems.

ACKNOWLEDGMENTS
The authors gratefully acknowledge valuable contributions and advice from P. Wolf, D. Boronson, H. Hemmati, A. Biswas, M. Krainak and M. Stevens on space lasercom links, S. Buchman, D. DeBra and F. Everitt on drag-free satellite systems, P. Enge and S. D'Amico on GNSS systems and POD performance.  G. Blewitt on GNSS-geodesy, C. Salomon on ACES and related, and Y. Bar-Sever on GRASP and related. This work is supported in part by the NASA - Fundamental Physics Program.